\documentclass[aps,prb,superscriptaddress,footinbib,twocolumn,10pt]{revtex4-2}

\usepackage{bm}
\usepackage{float}
\usepackage[colorlinks=true,allcolors=blue]{hyperref}
\usepackage{mhchem}
\usepackage{cleveref}
\usepackage{graphicx}
\usepackage{siunitx}

\crefname{figure}{Fig.}{Figs.}
\crefname{equation}{Eq.}{Eqs.}

\usepackage{etoolbox}
\patchcmd{\thebibliography}{\@openbib@code}{\@openbib@code\vspace{-1.5em}}{}{}

\begin{document}

\begin{abstract}
    Studies of electronic transport in width-restricted channels of \ce{PdCoO2} have recently revealed a novel `directional ballistic' regime, in which ballistic propagation of electrons on an anisotropic Fermi surface breaks the symmetries of bulk transport. Here we introduce a magnetic field to this regime, studying channels of \ce{PdCoO2} and \ce{PtCoO2} along two crystallographically distinct directions and over a wide range of widths. We observe magnetoresistance distinct from that in the bulk, with features strongly dependent on channel orientation and becoming more pronounced the narrower the channel. Comparison to semiclassical theory establishes that magnetoresistance arises from field-induced modification of boundary scattering, and helps connect features in the data with specific electronic trajectories. However, the role of bulk scattering in our measurements is yet to be fully understood. Our results demonstrate that finite-size magnetotransport is sensitive to the anisotropy of Fermi surface properties, motivating future work to fully understand and exploit this sensitivity.
\end{abstract}

\title{Directional ballistic magnetotransport in the delafossite metals \ce{PdCoO2} and \ce{PtCoO2}}
\author{Michal Moravec}
\affiliation{
    Max Planck Institute for Chemical Physics of Solids,
    Dresden, Germany
}
\affiliation{
    Scottish Universities Physics Alliance,
    School of Physics and Astronomy,
    University of St Andrews,
    St Andrews, UK
}
\author{Graham Baker}
\email{graham.baker@cpfs.mpg.de}
\affiliation{
    Max Planck Institute for Chemical Physics of Solids,
    Dresden, Germany
}
\author{Maja D. Bachmann}
\affiliation{
    Max Planck Institute for Chemical Physics of Solids,
    Dresden, Germany
}
\author{Aaron Sharpe}
\affiliation{
    Department of Physics,
    Stanford University,
    Stanford, CA, USA
}
\affiliation{
    Stanford Institute for Materials and Energy Sciences,
    SLAC National Accelerator Laboratory,
    Menlo Park, CA, USA
}
\author{Nabhanila Nandi}
\affiliation{
    Max Planck Institute for Chemical Physics of Solids,
    Dresden, Germany
}
\author{Arthur W. Barnard}
\affiliation{
    Department of Physics,
    University of Washington,
    Seattle, WA, USA
}
\affiliation{
    Department of Materials Science and Engineering,
    University of Washington,
    Seattle, WA, USA
}
\author{Carsten Putzke}
\affiliation{
    Max Planck Institute for the Structure and Dynamics of Matter,
    Hamburg, Germany
}
\author{Seunghyun Khim}
\affiliation{
    Max Planck Institute for Chemical Physics of Solids,
    Dresden, Germany
}
\author{Markus K{\"o}nig}
\affiliation{
    Max Planck Institute for Chemical Physics of Solids,
    Dresden, Germany
}
\author{David Goldhaber-Gordon}
\affiliation{
    Department of Physics,
    Stanford University,
    Stanford, CA, USA
}
\affiliation{
    Stanford Institute for Materials and Energy Sciences,
    SLAC National Accelerator Laboratory,
    Menlo Park, CA, USA
}
\author{Philip J.W. Moll}
\affiliation{
    Max Planck Institute for the Structure and Dynamics of Matter,
    Hamburg, Germany
}
\author{Andrew P. Mackenzie}
\email{andy.mackenzie@cpfs.mpg.de}
\affiliation{
    Max Planck Institute for Chemical Physics of Solids,
    Dresden, Germany
}
\affiliation{
    Scottish Universities Physics Alliance,
    School of Physics and Astronomy,
    University of St Andrews,
    St Andrews, UK
}

\maketitle

\section*{Introduction}

The delafossite metals \ce{PdCoO2}, \ce{PtCoO2}, and \ce{PdCrO2} \cite{Rogers1971} have become known as benchmarks for studying quasi two-dimensional electrical transport in the limit of extremely high conductivity \cite{Mackenzie2017}. Mean free paths ranging from several microns to over twenty microns have been observed in bulk crystals \cite{Hicks2012,Takatsu2013,Kikugawa2016,Nandi2018,Zhakina2023pnas,Zhang2024} due to extremely high crystal quality in the conducting Pd or Pt planes \cite{Sunko2020prx}, the screening of out-of-plane impurity potentials \cite{Zhang2024,Baker2024prx}, and anomalously low electron-phonon scattering \cite{Hicks2012,Seo2023,Yao2024}. Due to the long mean free paths, they are ideal for study of the de Haas-van Alphen effect (dHvA) \cite{Hicks2012,Ok2013,Hicks2015,Arnold2020}, and due to their quasi two-dimensionality they yield excellent angle-resolved photoemission spectroscopy (ARPES) data \cite{Noh2009,Sobota2013,Noh2014,Kushwaha2015,Sunko2017,Sunko2020sciadv,Mazzola2022}.

Together, dHvA and ARPES establish that \ce{PdCoO2} and \ce{PtCoO2} have nearly cylindrical Fermi surfaces with pronounced faceting that gives them almost hexagonal cross-sections. The triangular symmetry of the conducting planes means that the bulk in-plane resistivity is isotropic, but if transport is studied in samples restricted in one or more dimensions to length scales of order the mean free path, pronounced directional effects can be observed due to the bulk symmetry being broken by the shape of the sample. The delafossites are chemically suitable for focused ion beam (FIB) micro-structuring, which has enabled a series of experiments studying these directional effects as well as quantum coherence over length scales of over ten microns \cite{Bachmann2019natcomm,Putzke2020,Bachmann2022,McGuinness2021,Zhakina2023prb,Bachmann2019thesis}. Width restriction can also be achieved by tuning the skin depth in frequency-dependent electrodynamic studies \cite{Baker2024prx}, and the effect of Fermi surface faceting is the subject of active theoretical research \cite{Cook2019,Valentinis2023,Baker2023}.

\begin{figure*}[!ht]
    \includegraphics[width=0.67\textwidth]{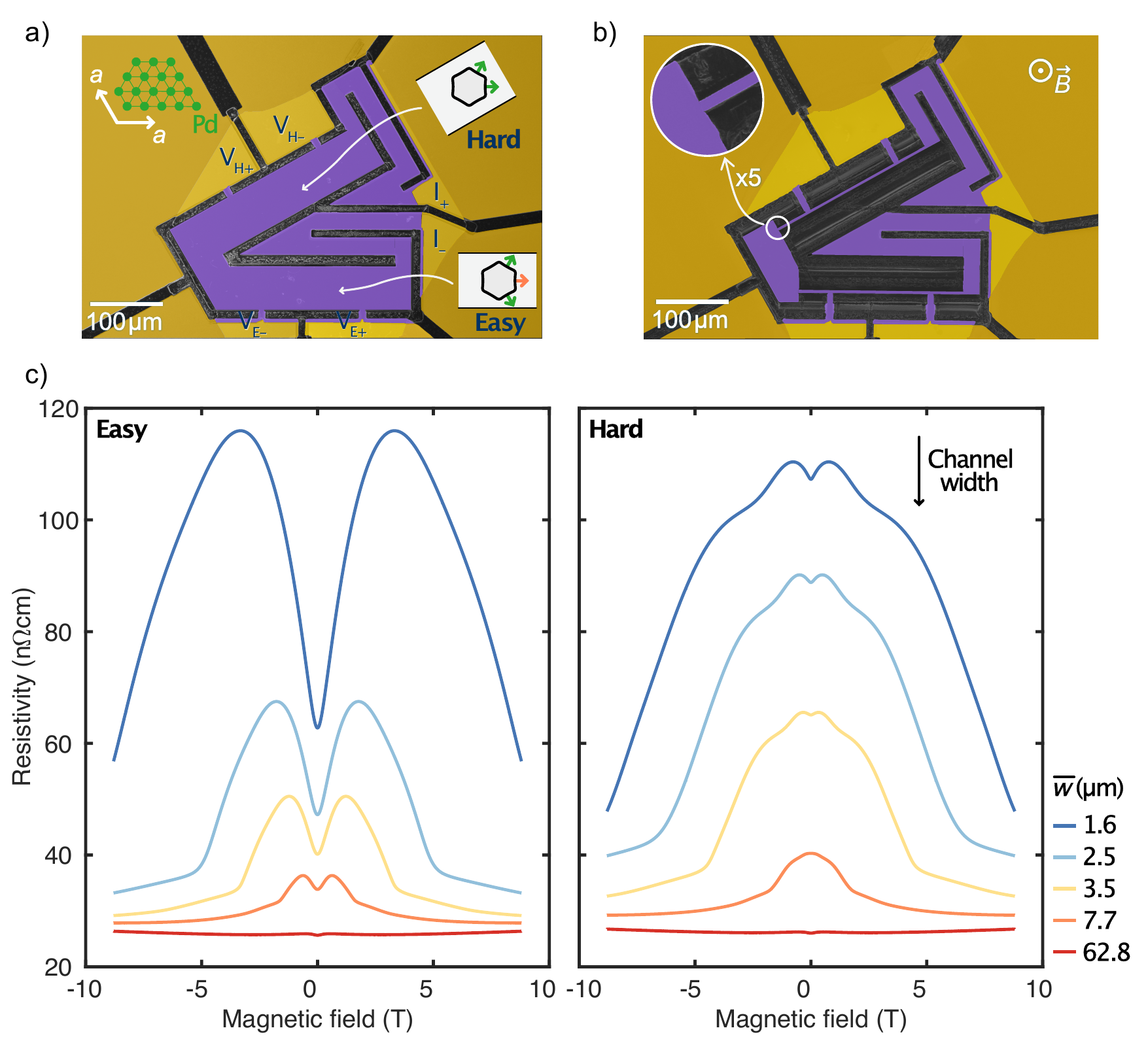}
    \caption{(a) False-colour scanning electron micrograph of the starting \ce{PdCoO2} device S1 for our experiment. Current is injected through gold contacts on the top of a single crystal of \ce{PdCoO2} and then flows through meander tracks to become homogeneous through the thickness of the sample before flowing through two conducting channels cut at 30$^{\circ}$ to one another. These are oriented relative to the Fermi surface facets in the so-called `hard' and `easy' directions previously identified in studies of directional transport in \ce{PdCoO2} in zero applied magnetic field \cite{Bachmann2022}, and are approximately \SI{63}{\micro\metre} wide. The voltage drops within the conducting channels along the easy and hard orientations were measured between contacts $V_{E+}$ and $V_{E-}$, and between $V_{H+}$ and $V_{H-}$ voltage contacts, respectively. (b) The same device after successive FIB narrowing to a channel width of \SI{3.5}{\micro\metre}. (c) Magnetoresistance data for five channel widths along the easy and hard directions. The legend shows the average of the channel widths in the two directions, $\bar{w}$.}
    \label{fig1}
\end{figure*}

Our goal in this paper is to report a comprehensive study of directional effects on the transverse magnetoresistance of width-restricted channels of \ce{PdCoO2} and \ce{PtCoO2} \footnote{We chose to focus only on magnetoresistance and not on Hall measurements because for the latter, it is challenging to distinguish intrinsic effects from those depending sensitively on the exact contact geometry \cite{Beenakker1991}. We verified that our magnetoresistance measurements did not vary with contact geometry.}. Size-restricted magnetotransport in this so-called MacDonald geometry \footnote{By ``transverse magnetoresistance in width-restricted channels'', we mean a geometry in which current, magnetic field, and size-restricted dimension are all along mutually perpendicular directions. This is as opposed to ``transverse magnetoresistance of thickness-restricted channels'', in which a magnetic field is perpendicular to current but parallel to the size-restricted dimension. The latter situation has also been stidued theoretically and experimentally \cite{Sondheimer1952}, but is conceptually and phenomenologically distinct from the present geometry.}  has received considerable theoretical attention \cite{MacDonald1950,Ditlefsen1966,Alekseev2016,Scaffidi2017,Alekseev2018,Alekseev2019,Holder2019,Raichev2020}, and has been studied experimentally in elemental metals \cite{MacDonald1950}, graphene \cite{Masubuchi2012}, and GaAs \cite{Raichev2020,Thornton1989,Gusev2018}. However, these works have focussed on materials in which the Fermi surface is well-approximated as isotropic. 
There has been some work on the effect of Fermi surface warping on ballistic magneto-transport, but only in more exotic geometries such as anti-dot lattices \cite{Zitzlsperger2003,Oka2019}, transverse electron focusing \cite{Bachmann2019natcomm}, and square junctions \cite{McGuinness2021}. 
While the MacDonald geometry was previously measured in \ce{PdCoO2} \cite{Moll2016}, at the time the importance of directional effects in \ce{PdCoO2} was not appreciated, so the channels were not aligned with the crystallographic axes. Since then, large directional effects have been reported in width-restricted channels of \ce{PdCoO2} \cite{Bachmann2022}, but only in zero magnetic field and not over a wide range of widths.

Here, we report on the effect of magnetic fields on transport in a single \ce{PdCoO2} device incorporating channels cut along the in-plane $a$ axis (previously established as the low-resistance, or ‘easy’ transport direction) and at \SI{30}{\degree} to it (the `hard' direction). Each channel was studied, then narrowed, in an iterative procedure that built up a large data set. In the main manuscript we concentrate on reporting the properties of this single device, but in supplementary information we show additional data sets from other \ce{PdCoO2} and \ce{PtCoO2} devices, demonstrating the consistency of the main trends that we identify in the data. We complement the experimental data with the results of Boltzmann calculations and Monte Carlo simulations for both idealized and realistic Fermi surface geometries. Our measurements and analysis demonstrate how this experimental arrangement can be used to study the anisotropy of Fermi surface geometry and scattering.

\section*{Results}

\begin{figure*}[!htbp]
    \includegraphics[width=0.99\textwidth]{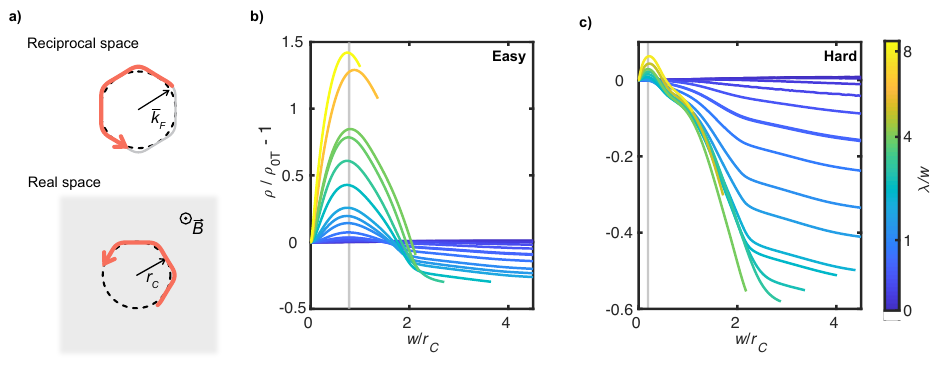}
    \caption{Magnetoresistance of \ce{PdCoO2} for a range of channel widths. Part (a) illustrates the semiclassical connection between the Fermi surface in reciprocal space and electron trajectories in real space. The hexagonal Fermi surface with average radius $\bar{k}_F$ translates into real-space cyclotron orbits of the same shape but rotated by \SI{90}{\degree}, with average radius $r_c$. Parts b) and c) show the complete data set for \ce{PdCoO2} S1 along the easy and hard directions, respectively. For each channel width, resistivity data are normalised to the value at zero field, with the ratio of the mean free path and channel width $\lambda/w$ indicated by the colour of the trace. Data are plotted against a second dimensionless variable, $w/r_{c}$. The vertical, grey lines indicate the average peak positions in magnetoresistance, at $w/r_{c}=0.78$ (easy) and 0.19 (hard).}
    \label{fig2}
\end{figure*}

A device (S1) used for our experiment is shown in the micrographs of \cref{fig1} (a) and (b). A single crystal of \ce{PdCoO2} \SI{1.8}{\micro\metre} thick is sculpted using a focused ion beam (FIB) into two conducting channels along the hard and easy directions identified in Ref. \cite{Bachmann2022}. Initially the width, $w$, of each channel was $\sim$ \SI{63}{\micro\metre}. Magnetoresistance was measured in a $^4\textrm{He}$ cryostat, and then the conducting channels were returned to the FIB and narrowed. This process was repeated 17 times down to \SI{0.75}{\micro\metre}, with the magnetoresistance measured along both directions at each step. To ensure the reliability of the experiment, considerable care must be taken to select crystal platelets of uniform thickness, and to maintain the width and separation of the voltage contacts on each channel. With the main crystal selection criterion being uniformity rather than the lowest attainable resistivity, the crystal shown in \cref{fig1} had a residual resistivity of \SI{26}{\nano\ohm\centi\metre}, approximately a factor of three higher than those attainable in the purest \ce{PdCoO2} \cite{Nandi2018,Bachmann2022,Moll2016}. 

Resistivity data from S1 are shown in \cref{fig1} (c). The easy-direction data are qualitatively similar to those reported in Ref. \cite{Moll2016} (for which the channel direction was not aligned precisely with a crystal axis) but those from the hard direction are qualitatively different. In both cases, the zero-field resistivity grows as the channel is narrowed, due to an increasing influence of boundary scattering, with the zero-field resistivity along the hard direction significantly larger than that along the easy direction for channels of similar width. There are then resistivity maxima as the field is increased, before the resistivity falls with increasing fields. The functional form of the magnetoresistance is clearly different for the two directions, but several features common to all the curves can be identified. Firstly, there is structure in the curves (the maxima and then pronounced kinks at higher fields) that scale approximately with field magnitude. Secondly, at high magnetic fields the hard and easy direction data look more similar to each other than at low fields, and seem to be tending to the value of the resistivity of the widest sample. 

Considering the semiclassical electron dynamics underlying our experiment suggests the relevant ingredients for understanding these trends. The semiclassical equations of motion for an electron in a band with dispersion $\varepsilon(\bm{k})$ are \footnote{Here $\bm{r}$ is real-space position, $\bm{k}$  is crystal momentum, $\bm{E}$ is electric field, $\bm{B}$ is magnetic field, $e$ is the electron charge, and $\hbar$ is the reduced Planck's constant.}
\begin{equation}\label{eq:vdot}
    \frac{d\bm{r}}{dt}
    \equiv\bm{v}
    =\frac{1}{\hbar}\nabla_{\bm{k}}\varepsilon(\bm{k})
\end{equation}
and
\begin{equation}\label{eq:kdot}
    \hbar\frac{d\bm{k}}{dt}=-e(\bm{E}+\bm{v}\times\bm{B}) .
\end{equation}
Since the Fermi surface is a constant-energy contour of $\varepsilon(\bm{k})$, \Cref{eq:vdot} dictates that Fermi velocity always points perpendicular to the Fermi surface, embedding the hexagonal Fermi surface shape into the electrons' motion.
In the presence of an out-of-plane magnetic field, as applied in our measurements, \cref{eq:kdot} shows that an electron will orbit around the Fermi surface. \Cref{eq:kdot} can furthermore be integrated to yield the real-space in-plane trajectory:
\begin{equation}
    \Delta\bm{r}_{\perp}(t)=\frac{\hbar}{eB}\Delta\bm{k}(t)\times\hat{c} .
\end{equation}
This means that the shape of the real-space cyclotron orbit is the same as that of the hexagonal Fermi surface, but rotated by \SI{90}{\degree} and scaled by the magnetic field according to $r_{c}=(\hbar/eB)\bar{k}_{F}$ \footnote{These statements apply to the in-plane motion, i.e. the projection of the real-space motion in the plane perpendicular to the $c$-axis magnetic field. The $c$-axis velocity is a constant of motion and the three-dimensional real-space path is in fact helical. However, the $c$-axis velocity is orders of magnitude smaller than the in-plane velocity, and so we do not consider the $c$-axis motion in our analysis}, where $r_{c}$ and $\bar{k}_{F}$ are orbitally-averaged real and reciprocal space scales, respectively.
These dynamics are illustrated in \cref{fig2}a.
In a real, finite lattice, scattering at the sample's boundaries and in its bulk acts to interrupt this semiclassical cyclotron motion. These considerations suggest three relevant length scales for our data: the cyclotron radius $r_{c}$, the channel's width $w$, and the mean free path for bulk scattering $\lambda$.

\begin{figure*}[!htbp]
    \includegraphics[width=0.8\textwidth]{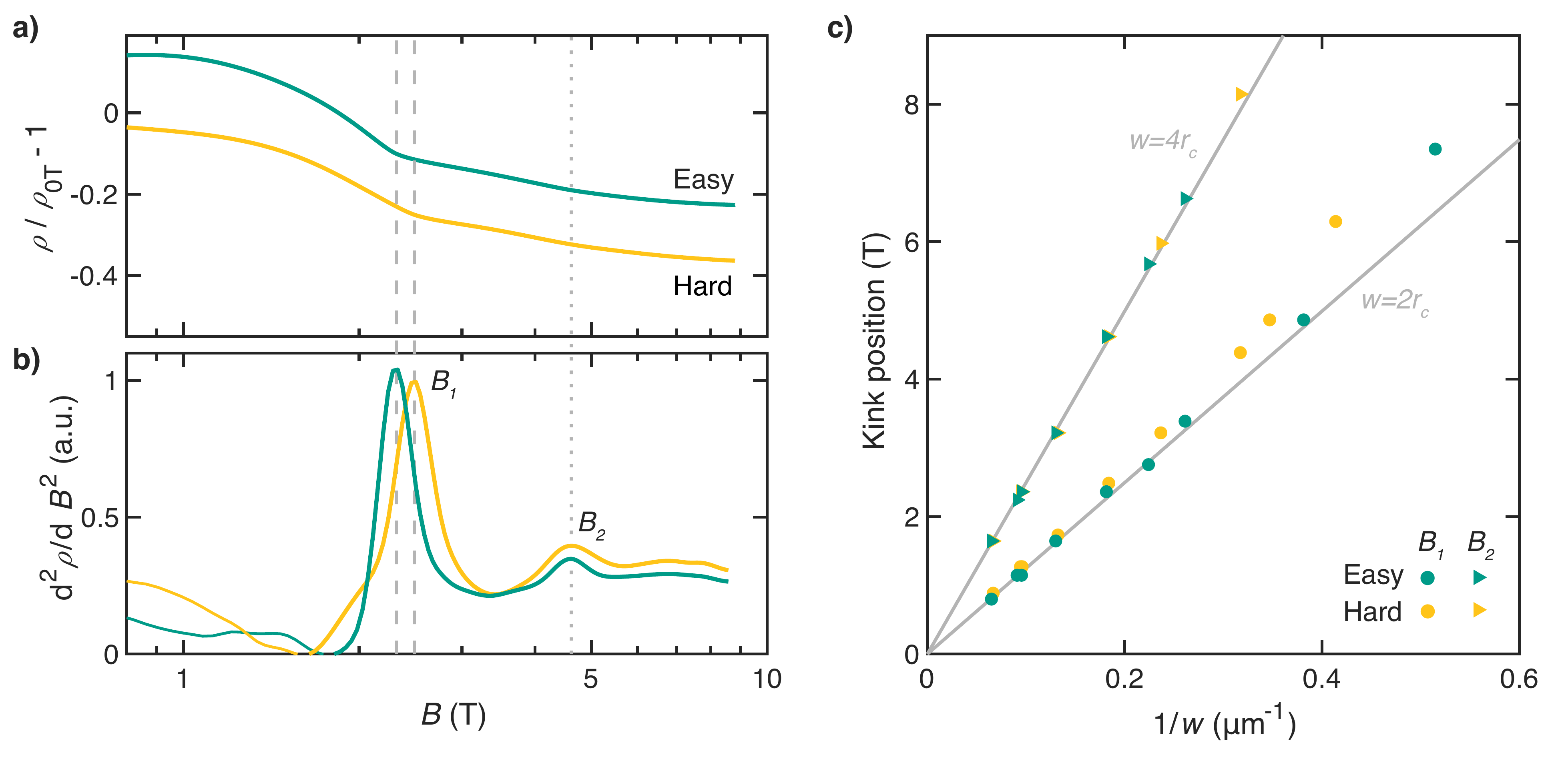}
    \caption{(a) Plotting the magnetoresistance against magnetic field on a semi-log scale reveals the existence of two kinks in the data at characteristic fields $B_{1}$ and $B_{2}$ for both easy and hard direction transport. (b) The exact position of the kinks was determined from local maxima in the second derivative of resistivity with respect to field. (c) Plotting $B_{1}$ and $B_{2}$ for samples with a wide range of widths against $1/w$ reveals that these kinks occur approximately to $w=2r_{c}$ and $w=4r_{c}$, respectively. For $B_{1}$, a small systematic difference between the easy and hard directions is resolved: the hard direction kink occurs at a slightly larger channel width than $2r_{c}$.}
    \label{fig3}
\end{figure*}

In order to examine the trends identified in the discussion of \cref{fig1}, we show in \cref{fig2} the magnetoresistance at each width $w$, defined relative to the zero-field resistivity at that width, plotted against $w/r_c$. Plotting magnetoresistance removes the effect of width-dependent boundary scattering at zero-field and instead emphasizes the relative influence of the magnetic field on boundary scattering. The colour of the traces represents the value of $\lambda/w$, emphasizing that the overall scale of the magnetoresistance has a clear systematic dependence on this dimensionless ratio. 

The main field-dependent features seen in \cref{fig1} do indeed occur at the same values of $w/r_c$, and the reduction of the resistivity in narrow channels from its zero-field value at sufficiently high $w/r_c$  is highlighted. For the easy direction, the peak in the magnetoresistance occurs at $w/r_c\approx0.6$, similar to a previous observation \cite{Moll2016}, while for the hard direction the overall maximum occurs at much smaller $w/r_c$, followed by a shoulder feature. In both orientations, a kink in the data is visible at $w/r_c\approx2$. 

To investigate the kink at $w/r_c\approx2$ in more detail, we plot data from a representative width of approximately \SI{5.5}{\micro\metre}  in \cref{fig3} on a semi-log plot. This reveals that we in fact resolve two kinks, the second at approximately twice the field of the first. Identifying the kinks via local maxima in $\partial^{2}\rho/\partial B^{2}$ (as described in the Supplementary Information) across samples with a range of widths for which they are accessible in the \SI{9}{\tesla} magnet used to study S1, we observe that they are indeed consistently observed at approximately $2r_c$ and $4r_c$ respectively, as shown in \cref{fig3}c). 

In addition to the data presented in \cref{fig1,fig2,fig3}, comprehensive measurements from an additional \ce{PdCoO2} sample as well as \ce{PtCoO2} sample are presented in the Supplementary Information. The data from all samples of a given material are fully consistent with one another, and the trends identified here for \ce{PdCoO2} are also seen in \ce{PtCoO2}. The comparison is discussed further in the Supplementary Information.

\section*{Discussion}

\begin{figure*}[!htbp]
    \includegraphics{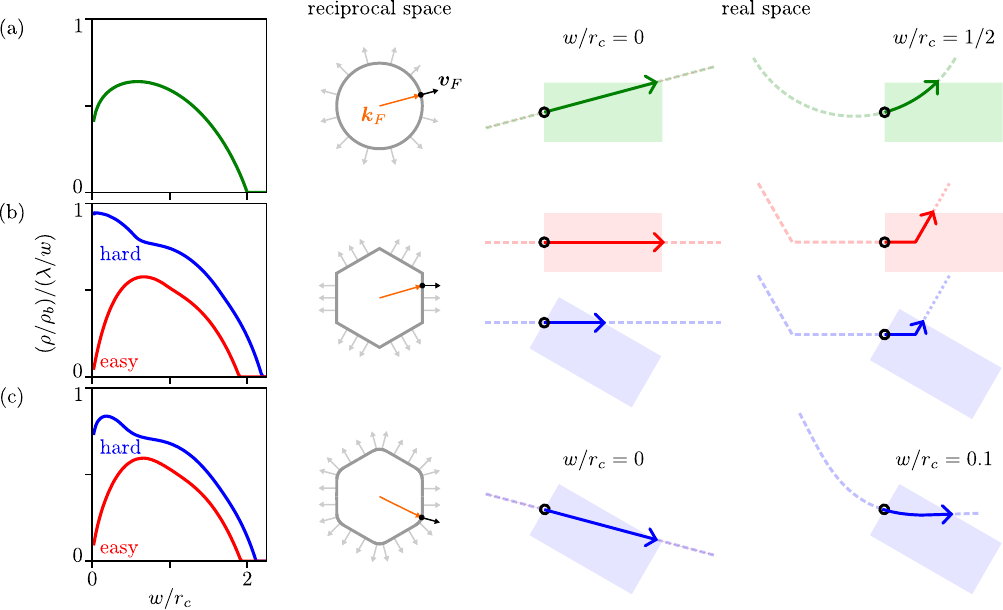}
    \caption{Boltzmann calculations and illustrations of low-field behaviour in the limit of no bulk scattering. We plot the dimensionless ratio $(\rho/\rho_{b})/(\lambda/w)$ where $\rho_{b}$ is the bulk resistivity in zero field. (a) For a circular Fermi surface, the resistivity shows a peak at $w/r_{c}\approx0.55$ as the magnetic field bends electrons towards the sample's edges. (b) For a hexagonal Fermi surface, at low magnetic fields with $0<w/r_{c}<1/2$, the magnetic field enhances edge scattering more strongly in the easy direction than in the hard direction. The easy orientation shows a large field-induced resistivity enhancement, whereas the resistivity in the hard direction is monotonically decreasing. (c) Moving to a realistic Fermi surface for \ce{PdCoO2} introduces an initial resistivity increase in the hard orientation at the lowest magnetic fields. This arises from states near the rounded corners of the Fermi surface, which avoid the edges at zero field but are bent towards them by relatively weak fields.}
    \label{fig4}
\end{figure*}

To set the context for our discussion, we summarize the set of experimental features, established in Figures \ref{fig2} \& \ref{fig3}, with which our analysis must contend. At low magnetic fields, the data show a rapid rise and then fall as magnetic field is increased. The form and magnitude of this behaviour differ significantly between the two channel orientations. For a given channel orientation, the position of the magnetoresistance peak appears to occur at a set value of $w/r_c$, and its magnitude increases as the ratio $\lambda/w$ is increased. At high fields, the resistivity decreases with increasing magnetic field, displaying two kinks at which the slope of this behaviour changes. The second kink appears to always occur at $w/r_c=4$, while the first kink occurs near $w/r_c=2$ but with a precise value that is systematically different between channel orientations. In all cases, the kinks become more pronounced with increasing $\lambda/w$.  

A theoretical model suitable for quantitatively analysing the full set of experimental features would need to incorporate Fermi surface anisotropy, scattering at the sample's boundaries, and scattering in the sample's bulk. To our knowledge, such a complete model does not currently exist. However, we believe that the experimental feature set can be understood at a qualitative level via a multi-pronged approach. This is enabled by the separable influences of the different length scales involved: the location of features in the magnetoresistance depends on $w/r_c$, while their magnitude depends on $\lambda/w$. 

Here is the approach that we will follow for the remainder of our discussion. To understand the $w/r_{c}$-dependent magnetoresistance features, we will make use of geometric models of field-dependent electronic trajectories. At low fields ($w/r_{c}<2$), we analyse the magnetoresistance peak with the help of Boltzmann calculations for a realistic Fermi surface geometry, but in the simplifying limit of no bulk scattering. (In the Supplementary Information, we also show the results of Monte Carlo simulations for the same set of assumptions.) At high fields ($w/r_{c}>2$), we analyse the magnetoresistance kinks with the help of simple geometric arguments about the phase space for different types of orbit. In both regimes, we then rely on qualitative arguments to explain how the magnitude of these features depends on bulk scattering, as quantified by the ratio $\lambda/w$.

\begin{figure*}
    \includegraphics{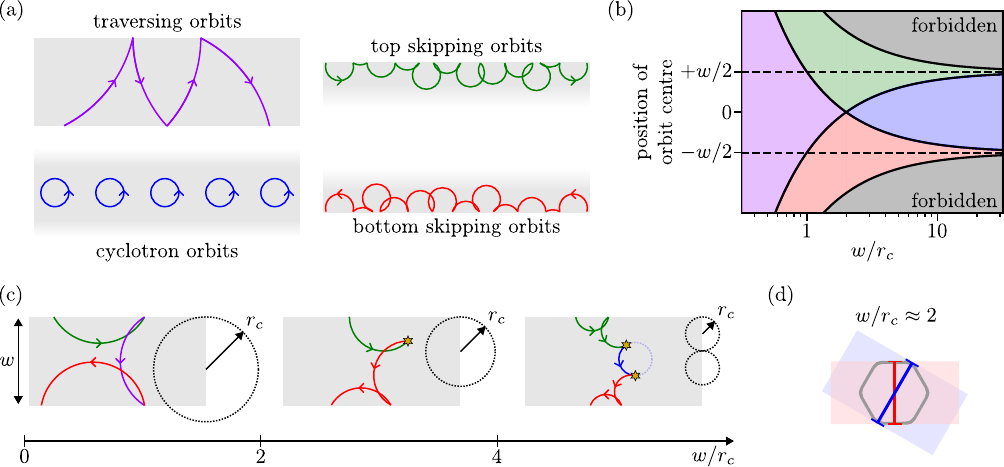}
    \caption{(a) Illustration of different orbit types. Orbits can be classified by whether they interact with one edge, both, or neither. For $w/r_{c}<2$, the sample supports traversing orbits, making contact with both edges. For $w/r_{c}>2$, the sample supports closed cyclotron orbits, making contact with neither edge. At all fields, the sample supports skipping orbits. Thee orbits start and end at the same edge, and travel in opposite directions at opposite edges. (b) Illustration of the phase space for each orbit type as a function of $w/r_{c}$. The plot indicates the possible positions of the orbit centre (the hypothetical centre point of the closed orbit) relative to the channel's boundaries, marked by dashed lines. (c) Depiction of routes to inter-edge scattering, which is necessary for width-dependent magnetoresistance. For $w/r_{c}<2$, no bulk scattering is required for inter-edge scattering. For $2<w/r_{c}<4$, at least one bulk scattering event is required. For $w/r_{c}>4$, at least two bulk scattering events are required. (d) For a realistic Fermi surface, the condition for the onset of closed cyclotron orbits is not exactly $w/r_{c}=2$ and depends on channel orientation.}
    \label{fig5}
\end{figure*}




The low field behaviour, shown in \cref{fig4}, is mainly characterized by resistivity enhancement from field-induced boundary scattering. The theoretically-expected behavior for an isotropic Fermi surface has previously been established \cite{Ditlefsen1966,Raichev2020}. Its main low-field feature is a peak at $w/r_c\approx0.55$ as electrons are bent towards the boundaries, which is reproduced by our Boltzmann calculation for a circular Fermi surface (\cref{fig4}a). However, it is clear that this model is insufficient for describing the direction-dependent changes in the qualitative form of the magnetoresistance seen in our data.

This motivates examining the influence of a nearly hexagonal Fermi surface geometry. The effect of Fermi surface geometry and orientation in \ce{PdCoO2} on boundary scattering at zero magnetic field was established in Ref. \cite{Bachmann2022}. In the hard orientation, electrons are always oriented toward the boundaries. In the easy orientation, a substantial fraction of electrons is oriented along the direction of the channel and avoids boundary scattering. 

Our data show a much larger field-induced resistivity enhancement in the easy direction than in the hard direction. This can be understood qualitatively by considering the difference between zero-field and low-field trajectories for a perfectly hexagonal Fermi surface, as shown in \cref{fig4}b. For the easy orientation, the trajectories that avoid boundary scattering at zero field are oriented toward the boundaries once even a small field is introduced. This creates a peak similar to that for a circular Fermi surface, but of even greater magnitude. On the contrary, for the hard orientation, a small field makes little difference as to whether trajectories reach the boundaries. This results in the absence of a low-field peak.

The above picture, based on an idealized hexagonal Fermi surface, misses another prominent feature of our data in the low-field regime: the small, narrow peak in the hard orientation visible for $w/r_{c}<1/2$. This peak is reproduced by a calculation using a realistic Fermi surface parameterization based on ARPES measurements \cite{Sunko2019}, as shown in \cref{fig4}c. This suggests that the peak occurs as electrons near the rounded corners of the Fermi surface are bent toward the boundaries by the magnetic field.

While our calculations do not account for the role of bulk scattering, our data show that it has a large effect. The peak positions appear to be determined by $w/r_c$, however \cref{fig2} demonstrates that the peak magnitudes are strongly dependent on $\lambda/w$. Since bulk scattering events interrupt the semiclassical electron trajectories, one would expect that the magnetic field would be more effective at inducing additional boundary scattering if there are fewer bulk scattering events. This is in line with the trend seen in the data. 

Interestingly, we have identified that the easy-orientation peak is dominated by electrons at the edges of the Fermi surface, whereas the hard-orientation peak is dominated by electrons at the corners of the Fermi surface. This offers the opportunity to separately study scattering at the edges and corners of the Fermi surface. Previous work on \ce{PdCoO2} has suggested that scattering at the edges and corners may differ, and that this difference may be instrumental in explaining several unconventional bulk transport properties \cite{Nandi2018}. If suitable theory is developed, our data may allow for a quantitative determination of any potential in-plane scattering anisotropy.

The features of the high-field regime, defined by $w/r_c>2$ and depicted in \cref{fig5}, can be understood by considering the possible types of orbits in a finite-width sample as a function of $w/r_c$ For simplicity, we consider a circular Fermi surface, though the same reasoning applies to the real Fermi surface of \ce{PdCoO2}. As depicted in \cref{fig5}a, one can delineate four types of orbits \cite{Beenakker1989}: (1) traversing orbits, touching both boundaries, (2) cyclotron orbits, touching neither boundary, and (3), (4) top and bottom skipping orbits, touching one boundary or the other. Simple geometry allows for constructing a phase diagram (\cref{fig5}b) for the possible locations of each orbit type relative to the channel. 

In the typical, hypothetical picture of a bulk magnetoresistance measurement, one ignores the presence of sample boundaries, and assumes that the bulk magnetoresistance arises only from cyclotron orbits. On the other hand, a size-dependent contribution to the magnetoresistance, scaling as $w/r_c$, must necessarily arise from trajectories interacting with both sample boundaries. Measurements in Ref. \cite{Nandi2018} show that bulk magnetoresistance is a weak effect in \ce{PdCoO2}—below 15\% at a temperature of 2 K and a magnetic field of 9 T. While the slight positive magnetoresistance visible in \cref{fig2} at high fields in our widest samples is likely attributable to a bulk contribution, the dominant features in our data scale as $w/r_c$  and are therefore attributable to a size-dependent contribution.

Therefore, to understand the size-dependent signal, we must consider trajectories interacting with both boundaries. At fields with $w/r_c<2$, traversing orbits connect the two boundaries, as shown in \cref{fig5}c (left). Once $w/r_c>2$, traversing orbits no longer exist. The only way for an electron to travel between boundaries is via a trajectory involving at least one bulk scattering event, transferring an electron between top and bottom skipping orbits (\cref{fig5}c, middle). Once $w/r_c>4$, a trajectory between top and bottom skipping orbits must involve an intermediate cyclotron orbit, and therefore requires at least two bulk scattering events (\cref{fig5}c, right). The change in the number of required bulk scattering events for inter-edge scattering at $w/r_c=2$ and $w/r_c=4$ are the origin of the $B_1$ and $B_2$ kinks identified in the data. 

If we relax our simplifying assumption of a circular Fermi surface, the most significant modification to the above picture is depicted in \cref{fig5}d. Our data showed a systematic difference in the position of the $B_1$ kink between the two orientations. Our calculations confirm that this a direct consequence of Fermi surface geometry: in one orientation, the kink position is set by the corner-to-corner dimension of the Fermi surface, while in the other, the kink position is set by the edge-to-edge dimension of the Fermi surface. 

While the discussion so far has focused on the ratio of width $w$ to cyclotron radius $r_c$, bulk scattering also plays a role. For a sample with $\lambda\gg w$, an electron will scatter only rarely as it traverses a channel, and so the bulk collisions required for a size-dependent signal for $w/r_c>2$ are rare. This suppresses the magnetoresistance in this regime, explaining the sharp kinks seen in our data in the narrower channels. For $\lambda<w$, an electron is likely to scatter as it traverses a channel, and it is not difficult to fulfill the requirement for a size-dependent signal. This is reflected in the absence of resolvable kinks in our data for the widest channels.

While we have discussed the effect of bulk scattering within the framework of momentum-relaxing scattering events that randomize an electron’s momentum, we cannot rule out the presence of small viscous corrections to the observed data. As described recently \cite{Baker2024prx,Baker2024adp}, such corrections are visible in microwave spectroscopy experiments on oriented \ce{PdCoO2} crystals. Their microscopic origin is thought to be small angle scattering from extended impurity potentials, a phenomenon which would also be present in the devices studied here. Searching for their effect on magnetotransport would likely require Boltzmann transport modelling including an applied magnetic field and a realistic parameterization of the \ce{PdCoO2} Fermi surface. Such a calculation presents significant technical difficulties, and to our knowledge no code to perform it currently exists. Until it does, it is unlikely to be possible to identify and quantify small viscous contributions to data such as those presented here.

It is interesting to note that for any absolute channel width, one would eventually expect a complete suppression of boundary-induced resistance at high enough magnetic field. This should occur for two reasons: (1) in the limit that $r_c\ll w$, the majority of electrons occupy cyclotron orbits, and (2) in the limit that $r_c\ll\lambda$, for those electrons remaining in skipping orbits, the bulk scattering necessary to undergo inter-edge scattering is greatly suppressed \cite{Bttiker1988}. Especially in materials with moderate and saturating bulk magnetoresistance---as in PdCoO$_2$, PtCoO$_2$ and many metals---this implies the opportunity for a sizable high-field decrease in resistivity for narrow channels with significant zero-field boundary scattering. 

Although the effects presented here occur at low temperature and high magnetic field, they nonetheless point to potential applications. Previous work has shown that the rate at which resistivity increases with decreasing channel width is slower in the easy orientation of PdCoO$_2$  and PtCoO$_2$ than in materials with isotropic Fermi surfaces \cite{Bachmann2022}, a desirable property in the search for new materials for high-conductivity interconnects in integrated circuits \cite{Kumar2022}. Building on this, the negative magnetoresistance we observe for $w/r_{c}>2$ provides a route to almost completely suppress boundary-induced resistivity, independent of channel orientation. Conversely, the large positive magnetoresistance that emerges for $w/r_{c}<2$ could enable applications in magnetic field sensing or switching. Although the magnetic field sensitivity is lower than that of existing sensors \cite{Daughton1999}, the present mechanism offers complementary possibilities because of the geometric tunability of the magnetoresistance.

An additional, broader implication of our measurements is that care is needed when interpreting magnetoresistance in highly-conducting materials. With increasing ability to work with smaller starting materials, there is opportunity for samples to inadvertently be in the quasi-ballistic regime \cite{Bachmann2022}. Given the widespread use of the bar geometry used here, care is needed in order not to accidentally conflate boundary-induced and intrinsic magnetoresistance. To complicate the issue, boundary-induced magnetoresistance may not follow the previously-established form for an idealized, isotropic Fermi surface. Instead, as demonstrated here, it can be strongly dependent on Fermi surface geometry.

In conclusion, we have presented a comprehensive data set describing directional finite-size magnetotransport in oriented conducting channels FIB sculpted from single crystals of delafossite metals. Our work illustrates the utility of finite-size magnetotransport. By tuning cyclotron radius relative to sample width, we were able to concretely relate features of our dataset to geometrical properties of the Fermi surface. By tuning both of these length scales relative to intrinsic scattering lengths, we were able to measure the effect of bulk scattering in significantly more detail than possible in a conventional transport measurement. At the moment, appropriate theory for quantitatively assessing the role of scattering has not yet been developed. We hope to motivate such theoretical work, enabling quantitative insights to be taken from our data set. 

\begin{acknowledgments}
    M.M. thanks Kent Shirer for insightful help with the micro-structure fabrication. This project was supported by the Max Planck Society and the European Research Council (ERC) under the European Union's Horizon 2020 research and innovation programme (MiTopMat, grant agreement no. 715730). M.D.B. acknowledges EPSRC for PhD studentship support through grant number EP/L015110/1. A.S. would like to thank Zack Gomez and Edwin Huang for helpful discussions and Tom Devereaux for letting us use his group cluster. Computational work was performed on the Sherlock cluster at Stanford University and on resources of the National Energy Research Scientific Computing Center, supported by DOE under contract DE-AC02-05CH11231. Analysis was supported by the US Department of Energy, Office of Science, Basic Energy Sciences, Materials Sciences and Engineering Division, under Contract DE-AC02-76SF00515. PJWM acknowledges financial support from the European Research Council (ERC) consolidator grant XBEND (No. 101080740).
\end{acknowledgments}

\appendix

\section*{Methods}

\subsection*{Single crystal growth and mounting}

Single crystals of \ce{PdCoO2} and \ce{PtCoO2} were grown as described elsewhere \cite{Kushwaha2015,Bachmann2019natcomm}. Crystals were mounted in Araldite epoxy on sapphire substrates. The structures were subsequently cured on a hot plate at $\SI{120}{\degreeCelsius}$ for $\SI{1}{\hour}$. Following curing, we deposited a $\SI{5}{\nano\meter}$ titanium binding layer and a $\SI{150}{\nano\meter}$ gold layer via sputtering before proceeding with focused ion beam (FIB) micro-structuring. The titanium/gold layer was removed from the central area to expose the crystal surface, where the device was subsequently defined with FIB.

\subsection*{Focused ion beam micro-structure fabrication}

We used the Ga-based FEI Helios NanoLab G3 CX FIB with a \SI{30}{\kilo\volt} accelerating voltage, following established FIB microstructuring procedure for delafossite metals \cite{Moll2016,Bachmann2019natcomm,Moll2018}. This procedure preserves the bulk of the crystal, as has been previously validated through the consistency of resistivity and quantum oscillations between bulk and FIB-structured samples \cite{Hicks2012,Moll2016,Nandi2018,Bachmann2019natcomm}. Simulations found that FIB structuring introduces a $\sim\SI{20}{\nano\metre}$ amorphized surface layer \cite{Moll2016}. For our narrowest channel of $\SI{0.7}{\micro\metre}$, the total sidewall damage ($\SI{20}{\nano\metre}$ on each side) represents less than $6\%$ of the channel width effectively reducing the channel width.

The FIB micro-structures contain current homogenization meanders ensuring current has dispersed through the thickness following a top contact injection. The large $c$-axis to $ab$-plane resistivity anisotropy causes preferential current flow through the top layers. Without proper homogenization, measurements would not reflect the true in-plane resistivity. The approximate length of the meander needed for current homogenization can be estimated considering current diffusion from the top layers as $L_{\text{meander}} \sim t\sqrt{\frac{\rho_c}{\rho_{ab}}}$.

\subsection*{Magnetotransport measurements}

Magnetotransport measurements were performed using a Physical Property Measurement System (PPMS) with samples mounted in 28-pin leadless chip carriers (LCCs) housed within a custom-designed insert. Initial characterization was conducted between $\SI{5}{\kelvin}$ and $\SI{300}{\kelvin}$ under magnetic fields ranging from $\SI{-9}{\tesla}$ to $\SI{9}{\tesla}$.

Electrical transport measurements utilized a bespoke common mode rejection current source coupled with a Synktek MCL1-540 lock-in amplifier operating at $\SI{177.8}{\hertz}$. Excitation currents ranged between $\SI{0.1}{\milli\ampere}$ and $\SI{9}{\milli\ampere}$. Before each measurement, we verified current-voltage characteristics and found no non-linearities. Common mode voltage rejection was confirmed by reversing voltage contact polarity, with measured values remaining unchanged within 1 part in 100.

\subsection*{Processing of magnetoresistance data}

The raw voltage values $V$ were converted to the resistivity $\rho$ based on the dimensions of the sample, width $w$, thickness $t$ and average contact separation $L$, and the applied current $I$ through the relation $\rho = \frac{V}{I} \frac{wt}{L}$. This resistivity was subsequently interpolated, averaged over the decreasing and increasing magnetic field sweeps, and symmetrized. To determine the kink positions, as illustrated in Fig.~\ref{fig3}, the data were grouped into 177 logarithmically-spaced bins between \SI{0.001}{\tesla} and \SI{8.8}{\tesla} to mitigate the effect of noise before calculating the second derivative of the resistivity ($d^2\rho/dB^2$). Finally, the plotted traces were normalized to the unit range via the transformation $(y - y_{\mathrm{min}})/(y_{\mathrm{max}} - y_{\mathrm{min}})$, where $y_{\text{min}}$ and $y_{\text{max}}$ are the minimum and maximum values of a given trace. This normalization emphasizes the field-dependent evolution of the features independent of the absolute magnitude of the derivative.

\subsection*{Determination of bulk mean free path and average cyclotron radius}

The bulk momentum-relaxing mean free path $\lambda$ was determined from the in-plane resistivity $\rho_{\text{bulk}}$ measured at $\SI{5}{\kelvin}$ in the widest channels. We obtained $\rho_{\text{bulk}} = \SI{25.8}{\nano\ohm\centi\metre}$ for \ce{PdCoO2} and $\SI{31.8}{\nano\ohm\centi\metre}$ for \ce{PtCoO2} by averaging the zero-field resistivity values from both easy and hard crystallographic orientations. The calculation used the standard two-dimensional expression, \mbox{$\lambda = \frac{e^2 \bar{k}_F}{h \, d} \frac{1}{\rho_{\text{bulk}}}$} \cite{Ziman1964}, where $e$ is the electron charge, $h$ is the Planck constant, $d$ is the interlayer spacing ($c/3$, where $c = \SI{17.743}{\angstrom}$ for \ce{PdCoO2} and $\SI{17.808}{\angstrom}$ for \ce{PtCoO2} is the respective $c$-axis lattice constant), $\SI{0.9518}{\per\angstrom}$ for \ce{PdCoO2} and $\SI{0.9542}{\per\angstrom}$ for \ce{PtCoO2} are the angle averaged Fermi momenta $\bar{k}_F \equiv \frac{1}{2\pi} \int_0^{2\pi} d\theta \, k_F(\theta)$ based on parameters from Ref.  \cite{Sunko2019}. This averaged value was also used to determine the cyclotron radius $r_c$ used throughout the main text.

\subsection*{Iterative channel narrowing}

Following initial measurements of the micro-structure, an iterative refinement process was employed where samples were transferred between the FIB system for channel narrowing and the PPMS for transport characterization. Measurement reliability throughout this process was confirmed through consistent room-temperature resistivity values. This FIB processing and PPMS re-measurement cycle was repeated 17 times for \ce{PdCoO2} S1 (resulting in 18 distinct channel width measurements), once for \ce{PdCoO2} S2 (2 measurements), and 15 times for \ce{PtCoO2} S1 (16 measurements). Channels were narrowed in both crystallographic orientations during each FIB processing step.

To achieve precise geometry of the narrowed channels and accurate width measurement, careful alignment and a uniform cross-section were essential. A cross-shaped marker aligned along the easy direction helped ensure the channel edges were parallel within $\SI{0.1}{\degree}$. Narrowing was performed using beam currents in the range $\SI{0.17}{\nano\ampere}$ to $\SI{0.79}{\nano\ampere}$ to minimize beam tail damage and edge rounding. For channels narrower than $\SI{5}{\micro\meter}$, the channels were narrowed from a different channel side in each iteration because the width reduction became typically sub-micrometer. The channel width was determined by measuring at the widest point ($w_\textrm{outer}$) and undamaged top surface width $w_{\text{inner}}$. We then calculated a weighted average $w = (w_{\text{inner}} + 2w_{\text{outer}})/3$ to account for the rounded edges resulting from beam tail effects.



\bibliography{main.bib}

@article{Alekseev2016,
   author = {P S Alekseev},
   doi = {10.1103/PhysRevLett.117.166601},
   journal = {Physical Review Letters},
   pages = {166601},
   title = {Negative Magnetoresistance in Viscous Flow of Two-Dimensional Electrons},
   volume = {117},
   year = {2016}
}

@article{Alekseev2018,
   author = {P. S. Alekseev and M. A. Semina},
   doi = {10.1103/PhysRevB.98.165412},
   issn = {24699969},
   journal = {Physical Review B},
   pages = {165412},
   title = {Ballistic flow of two-dimensional interacting electrons},
   volume = {98},
   year = {2018}
}

@article{Alekseev2019,
   author = {P. S. Alekseev and M. A. Semina},
   doi = {10.1103/PhysRevB.100.125419},
   issn = {24699969},
   journal = {Physical Review B},
   pages = {125419},
   title = {Hall effect in a ballistic flow of two-dimensional interacting particles},
   volume = {100},
   url = {https://doi.org/10.1103/PhysRevB.100.125419},
   year = {2019}
}

@article{Arnold2020,
   author = {F. Arnold and M. Naumann and H. Rosner and N. Kikugawa and D. Graf and L. Balicas and T. Terashima and S. Uji and H. Takatsu and S. Khim and A. P. Mackenzie and E. Hassinger},
   doi = {10.1103/PhysRevB.101.195101},
   issn = {24699969},
   journal = {Physical Review B},
   pages = {195101},
   title = {Fermi surface of \ce{PtCoO2} from quantum oscillations and electronic structure calculations},
   volume = {101},
   year = {2020}
}

@article{Bachmann2019natcomm,
   author = {Maja D Bachmann and Aaron L Sharpe and Arthur W Barnard and Carsten Putzke and Markus König and Seunghyun Khim and David Goldhaber-Gordon and Andrew P Mackenzie and Philip J W Moll},
   doi = {10.1038/s41467-019-13020-9},
   journal = {Nature Communications},
   pages = {5081},
   title = {Super-geometric electron focusing on the hexagonal {F}ermi surface of \ce{PdCoO2}},
   volume = {10},
   year = {2019}
}

@book{Bachmann2019thesis,
   author = {Maja Deborah Bachmann},
   isbn = {9783030513610},
   title = {Manipulating Anisotropic Transport and Superconductivity by Focused Ion Beam Microstructuring},
   year = {2019}
}

@article{Bachmann2022,
   author = {Maja D. Bachmann and Aaron L. Sharpe and Graham Baker and Arthur W. Barnard and Carsten Putzke and Thomas Scaffidi and Nabhanila Nandi and P H Mcguinness and Elina Zhakina and Michal Moravec and Seunghyun Khim and Markus König and David Goldhaber-Gordon and Andrew P. Mackenzie and Philip J. W. Moll},
   doi = {10.1038/s41567-022-01570-7},
   journal = {Nature Physics},
   pages = {819},
   title = {Directional ballistic transport in the two-dimensional metal \ce{PdCoO2}},
   volume = {18},
   year = {2022}
}

@article{Baker2023,
   author = {G Baker and D Valentinis and A. P. Mackenzie},
   doi = {10.1063/10.0022360},
   journal = {Low Temperature Physics},
   pages = {1338},
   title = {On non-local electrical transport in anisotropic metals},
   volume = {49},
   year = {2023}
}

@article{Baker2024prx,
   author = {Graham Baker and Timothy W. Branch and J. S. Bobowski and James Day and Davide Valentinis and Mohamed Oudah and Philippa McGuinness and Seunghyun Khim and Piotr Surówka and Yoshiteru Maeno and Roderich Moessner and Jörg Schmalian and Andrew P. Mackenzie and D. A. Bonn},
   doi = {10.1103/PhysRevX.14.011018},
   journal = {Physical Review X},
   pages = {011018},
   title = {Nonlocal Electrodynamics in Ultrapure \ce{PdCoO2}},
   volume = {14},
   year = {2024}
}

@article{Baker2024adp,
   author = {Graham Baker and Michal Moravec and Andrew P. Mackenzie},
   doi = {10.1002/andp.202400087},
   issn = {15213889},
   journal = {Annalen der Physik},
   pages = {2400087},
   title = {A Perspective on Non-Local Electronic Transport in Metals: Viscous, Ballistic, and Beyond},
   volume = {536},
   year = {2024}
}

@article{Beenakker1989,
   author = {C W J Beenakker and H van Houten and B J van Wees},
   doi = {10.1016/0749-6036(89)90081-5},
   journal = {Superlattices and Microstructures},
   pages = {127-32},
   title = {Skipping orbits, traversing trajectories, and quantum ballistic transport in microstructures},
   volume = {5},
   year = {1989}
}

@article{Bttiker1988,
   author = {M. Büttiker},
   doi = {10.1103/PhysRevB.38.9375},
   issn = {01631829},
   journal = {Physical Review B},
   pages = {9375},
   title = {Absence of backscattering in the quantum {H}all effect in multiprobe conductors},
   volume = {38},
   year = {1988}
}

@article{Cook2019,
   author = {Caleb Q Cook and Andrew Lucas},
   doi = {10.1103/PhysRevB.99.235148},
   journal = {Physical Review B},
   pages = {235148},
   title = {Electron hydrodynamics with a polygonal {F}ermi surface},
   volume = {99},
   year = {2019}
}

@article{Ditlefsen1966,
   author = {E. Ditlefsen and J. Lothe},
   doi = {10.1080/14786436608211970},
   journal = {Philosophical Magazine},
   pages = {759},
   title = {Theory of size effects in electrical conductivity},
   volume = {14},
   year = {1966}
}

@article{Gusev2018,
   author = {G. M. Gusev and A. D. Levin and E. V. Levinson and A. K. Bakarov},
   doi = {10.1103/PhysRevB.98.161303},
   issn = {24699969},
   journal = {Physical Review B},
   pages = {161303(R)},
   title = {Viscous transport and {H}all viscosity in a two-dimensional electron system},
   volume = {98},
   year = {2018}
}

@article{Hicks2012,
   author = {Clifford W Hicks and Alexandra S Gibbs and Andrew P Mackenzie and Hiroshi Takatsu and Yoshiteru Maeno and Edward A Yelland},
   doi = {10.1103/PhysRevLett.109.116401},
   journal = {Physical Review Letters},
   pages = {116401},
   title = {Quantum Oscillations and High Carrier Mobility in the Delafossite \ce{PdCoO2}},
   volume = {109},
   year = {2012}
}

@article{Hicks2015,
   author = {Clifford W. Hicks and Alexandra S. Gibbs and Lishan Zhao and Pallavi Kushwaha and Horst Borrmann and Andrew P. Mackenzie and Hiroshi Takatsu and Shingo Yonezawa and Yoshiteru Maeno and Edward A. Yelland},
   doi = {10.1103/PhysRevB.92.014425},
   issn = {1550235X},
   journal = {Physical Review B},
   pages = {014425},
   title = {Quantum oscillations and magnetic reconstruction in the delafossite \ce{PdCrO2}},
   volume = {92},
   year = {2015}
}

@article{Holder2019,
   author = {Tobias Holder and Raquel Queiroz and Thomas Scaffidi and Navot Silberstein and Asaf Rozen and Joseph A. Sulpizio and Lior Ella and Shahal Ilani and Ady Stern},
   doi = {10.1103/PhysRevB.100.245305},
   journal = {Physical Review B},
   pages = {245305},
   title = {Ballistic and hydrodynamic magnetotransport in narrow channels},
   volume = {100},
   year = {2019}
}

@article{Kikugawa2016,
   author = {N Kikugawa and P Goswami and A Kiswandhi and E S Choi and D Graf and R E Baumbach and J S Brooks and K Sugii and Y Iida and M Nishio and S Uji and T Terashima and P M C Rourke and N E Hussey and H Takatsu and S Yonezawa and Y Maeno and L Balicas},
   doi = {10.1038/ncomms10903},
   journal = {Nature Communications},
   pages = {10903},
   title = {Interplanar coupling-dependent magnetoresistivity in high-purity layered metals},
   volume = {7},
   url = {www.nature.com/naturecommunications},
   year = {2016}
}

@article{Kushwaha2015,
   author = {Pallavi Kushwaha and Veronika Sunko and Philip J.W. Moll and Lewis Bawden and Jonathon M. Riley and Nabhanila Nandi and Helge Rosner and Marcus P. Schmidt and Frank Arnold and Elena Hassinger and Timur K. Kim and Moritz Hoesch and Andrew P. Mackenzie and Phil D.C. King},
   doi = {10.1126/sciadv.1500692},
   journal = {Science Advances},
   pages = {1500692},
   title = {Nearly free electrons in a 5d delafossite oxide metal},
   volume = {1},
   year = {2015}
}

@article{MacDonald1950,
   author = {DKC MacDonald and K Sarginson},
   doi = {10.1098/rspa.1950.0136},
   issn = {0080-4630},
   journal = {Proceedings of the Royal Society A},
   pages = {223},
   title = {Size effect variation of the electrical conductivity of metals},
   volume = {203},
   year = {1950}
}

@article{Mackenzie2017,
   author = {Andrew P. Mackenzie},
   doi = {10.1088/1361-6633/aa50e5},
   journal = {Reports on Progress in Physics},
   pages = {32501},
   title = {The properties of ultrapure delafossite metals},
   volume = {80},
   url = {https://iopscience.iop.org/article/10.1088/1361-6633/aa50e5/pdf},
   year = {2017}
}

@article{Masubuchi2012,
   author = {Satoru Masubuchi and Kazuyuki Iguchi and Takehiro Yamaguchi and Masahiro Onuki and Miho Arai and Kenji Watanabe and Takashi Taniguchi and Tomoki MacHida},
   doi = {10.1103/PhysRevLett.109.036601},
   issn = {00319007},
   number = {3},
   journal = {Physical Review Letters},
   pages = {1-5},
   title = {Boundary scattering in ballistic graphene},
   volume = {109},
   year = {2012}
}

@article{McGuinness2021,
   author = {Philippa H. McGuinness and Elina Zhakina and Markus König and Maja D. Bachmann and Carsten Putzke and Philip J. W. Moll and Seunghyun Khim and Andrew P. Mackenzie},
   doi = {10.1073/PNAS.2113185118},
   journal = {Proceedings of the National Academy of Sciences of the United States of America},
   pages = {2113185118},
   title = {Low-symmetry nonlocal transport in microstructured squares of delafossite metals},
   volume = {118},
   year = {2021}
}

@article{Moll2016,
   author = {Philip J W Moll and Pallavi Kushwaha and Nabhanila Nandi and Burkhard Schmidt and Andrew P Mackenzie},
   doi = {10.1126/science.aac8385},
   issue = {6277},
   journal = {Science},
   pages = {1061-4},
   title = {Evidence for hydrodynamic electron flow in \ce{PdCoO2}},
   volume = {351},
   year = {2016}
}

@article{Moll2018,
   author = {Philip J.W. Moll},
   doi = {10.1146/annurev-conmatphys-033117-054021},
   issn = {19475462},
   journal = {Annual Review of Condensed Matter Physics},
   pages = {147},
   title = {Focused Ion Beam Microstructuring of Quantum Matter},
   volume = {9},
   year = {2018}
}

@article{Nandi2018,
   author = {Nabhanila Nandi and Thomas Scaffidi and Pallavi Kushwaha and Seunghyun Khim and Mark E Barber and Veronika Sunko and Federico Mazzola and Philip D C King and Helge Rosner and Philip J W Moll and Markus König and Joel E Moore and Sean Hartnoll and Andrew P Mackenzie},
   doi = {10.1038/s41535-018-0138-8},
   journal = {npj Quantum Materials},
   pages = {66},
   title = {Unconventional magneto-transport in ultrapure \ce{PdCoO2} and \ce{PtCoO2}},
   volume = {3},
   url = {https://doi.org/10.1038/s41535-018-0138-8},
   year = {2018}
}

@article{Noh2009,
   author = {Han-Jin Noh and Jinwon Jeong and Jinhwan Jeong and En-Jin Cho and Sung Baek Kim and Kyoo Kim and B I Min and Hyeong-Do Kim},
   doi = {10.1103/PhysRevLett.102.256404},
   journal = {Physical Review Letters},
   pages = {256404},
   title = {Anisotropic Electric Conductivity of Delafossite \ce{PdCoO2} Studied by Angle-Resolved Photoemission Spectroscopy},
   volume = {102},
   year = {2009}
}

@article{Noh2014,
   author = {Han Jin Noh and Jinwon Jeong and Bin Chang and Dahee Jeong and Hyun Sook Moon and En Jin Cho and Jong Mok Ok and Jun Sung Kim and Kyoo Kim and B. I. Min and Han Koo Lee and Jae Young Kim and Byeong Gyu Park and Hyeong Do Kim and Seongsu Lee},
   doi = {10.1038/srep03680},
   issn = {20452322},
   journal = {Scientific Reports},
   pages = {3680},
   title = {Direct Observation of Localized Spin Antiferromagnetic Transition in \ce{PdCrO2} by Angle-Resolved Photoemission Spectroscopy},
   volume = {4},
   year = {2014}
}

@article{Ok2013,
   author = {Jong Mok Ok and Y. J. Jo and Kyoo Kim and T. Shishidou and E. S. Choi and Han Jin Noh and T. Oguchi and B. I. Min and Jun Sung Kim},
   doi = {10.1103/PhysRevLett.111.176405},
   issn = {00319007},
   journal = {Physical Review Letters},
   pages = {176405},
   title = {Quantum oscillations of the metallic triangular-lattice antiferromagnet \ce{PdCrO2}},
   volume = {111},
   year = {2013}
}

@article{Putzke2020,
   author = {Carsten Putzke and Maja D. Bachmann and Philippa McGuinness and Elina Zhakina and Veronika Sunko and Marcin Konczykowski and Takashi Oka and Roderich Moessner and Ady Stern and Markus König and Seunghyun Khim and Andrew P. Mackenzie and Philip J.W. Moll},
   doi = {10.1126/SCIENCE.AAY8413/FORMAT/PDF},
   issn = {10959203},
   issue = {6496},
   journal = {Science},
   month = {6},
   pages = {1234--1238},
   pmid = {32527829},
   publisher = {American Association for the Advancement of Science},
   title = {{$h/e$} oscillations in interlayer transport of delafossites},
   volume = {368},
   url = {https://www.science.org},
   year = {2020}
}

@article{Raichev2020,
   author = {O. E. Raichev and G. M. Gusev and A. D. Levin and A. K. Bakarov},
   doi = {10.1103/PhysRevB.101.235314},
   journal = {Physical Review B},
   pages = {235314},
   publisher = {American Physical Society},
   title = {Manifestations of classical size effect and electronic viscosity in the magnetoresistance of narrow two-dimensional conductors: Theory and experiment},
   volume = {101},
   year = {2020}
}

@article{Rogers1971,
   author = {D. B. Rogers and R. D. Shannon and C. T. Prewitt and J. L. Gillson},
   doi = {10.1021/ic50098a013},
   issn = {1520510X},
   journal = {Inorganic Chemistry},
   pages = {723},
   title = {Chemistry of Noble Metal Oxides. III. Electrical Transport Properties and Crystal Chemistry of \ce{ABO2} Compounds with the Delafossite Structure},
   volume = {10},
   year = {1971}
}

@article{Scaffidi2017,
   author = {Thomas Scaffidi and Nabhanila Nandi and Burkhard Schmidt and Andrew P. Mackenzie and Joel E. Moore},
   doi = {10.1103/PhysRevLett.118.226601},
   journal = {Physical Review Letters},
   pages = {226601},
   title = {Hydrodynamic Electron Flow and {H}all Viscosity},
   volume = {118},
   year = {2017}
}

@article{Sobota2013,
   author = {J. A. Sobota and K. Kim and H. Takatsu and M. Hashimoto and S. K. Mo and Z. Hussain and T. Oguchi and T. Shishidou and Y. Maeno and B. I. Min and Z. X. Shen},
   doi = {10.1103/PhysRevB.88.125109},
   issn = {10980121},
   journal = {Physical Review B},
   pages = {125109},
   title = {Electronic structure of the metallic antiferromagnet \ce{PdCrO2} measured by angle-resolved photoemission spectroscopy},
   volume = {88},
   year = {2013}
}

@article{Sondheimer1952,
   author = {E.H. Sondheimer},
   doi = {10.1080/00018735200101151},
   journal = {Advances in Physics},
   pages = {1-42},
   title = {The mean free path of electrons in metals},
   volume = {1},
   year = {1952}
}

@phdthesis{Sunko2019,
   author = {Veronika Sunko},
   doi = {10.1007/978-3-030-31087-5_4},
   school = {University of St Andrews},
   title = {Angle Resolved Photoemission Spectroscopy of Delafossite Metals},
   year = {2019}
}

@article{Sunko2020prx,
   author = {V Sunko and P H McGuinness and C S Chang and E Zhakina and S Khim and C E Dreyer and M Konczykowski and H Borrmann and P J W Moll and M König and D A Muller and A P Mackenzie},
   doi = {10.1103/PhysRevX.10.021018},
   journal = {Physical Review X},
   pages = {021018},
   title = {Controlled Introduction of Defects to Delafossite Metals by Electron Irradiation},
   volume = {10},
   year = {2020}
}

@article{Sunko2017,
   author = {Veronika Sunko and H. Rosner and P. Kushwaha and S. Khim and F. Mazzola and L. Bawden and O. J. Clark and J. M. Riley and D. Kasinathan and M. W. Haverkort and T. K. Kim and M. Hoesch and J. Fujii and I. Vobornik and A. P. Mackenzie and P. D.C. King},
   doi = {10.1038/nature23898},
   journal = {Nature},
   pages = {492},
   title = {Maximal {R}ashba-like spin splitting via kinetic-energy-coupled inversion-symmetry breaking},
   volume = {549},
   year = {2017}
}

@article{Sunko2020sciadv,
   author = {V. Sunko and F. Mazzola and S. Kitamura and S. Khim and P. Kushwaha and O. J. Clark and M. D. Watson and I. Marković and D. Biswas and L. Pourovskii and T. K. Kim and T. L. Lee and P. K. Thakur and H. Rosner and A. Georges and R. Moessner and T. Oka and A. P. Mackenzie and P. D.C. King},
   doi = {10.1126/sciadv.aaz0611},
   issn = {23752548},
   journal = {Science Advances},
   pages = {eaaz06117},
   pmid = {32128385},
   title = {Probing spin correlations using angle-resolved photoemission in a coupled metallic/{M}ott insulator system},
   volume = {6},
   year = {2020}
}

@article{Takatsu2013,
   author = {Hiroshi Takatsu and Jun J Ishikawa and Shingo Yonezawa and Harukazu Yoshino and Tatsuya Shishidou and Tamio Oguchi and Keizo Murata and Yoshiteru Maeno},
   doi = {10.1103/PhysRevLett.111.056601},
   journal = {Physical Review Letters},
   pages = {056601},
   title = {Extremely Large Magnetoresistance in the Nonmagnetic Metal \ce{PdCoO2}},
   volume = {111},
   year = {2013}
}

@article{Thornton1989,
   author = {TJ Thornton and ML Roukes and A Scherer and BP Van de Gaag},
   journal = {Physical Review Letters},
   pages = {2128},
   title = {Boundary Scattering in Quantum Wires},
   volume = {63},
   year = {1989}
}

@article{Valentinis2023,
   author = {Davide Valentinis and Graham Baker and Douglas A. Bonn and Jörg Schmalian},
   doi = {10.1103/PhysRevResearch.5.013212},
   journal = {Physical Review Research},
   pages = {013212},
   publisher = {American Physical Society},
   title = {Kinetic theory of the non-local electrodynamic response in anisotropic metals: skin effect in {2D} systems},
   volume = {5},
   year = {2023}
}

@article{Zhakina2023prb,
   author = {Elina Zhakina and Philippa H. McGuinness and Markus König and Romain Grasset and Maja D. Bachmann and Seunghyun Khim and Carsten Putzke and Philip J. W. Moll and Marcin Konczykowski and Andrew P. Mackenzie},
   doi = {10.1103/physrevb.107.094203},
   journal = {Physical Review B},
   pages = {094203},
   title = {Crossing the ballistic-ohmic transition via high energy electron irradiation},
   volume = {107},
   year = {2023}
}

@article{Zhakina2023pnas,
   author = {Elina Zhakina and Ramzy Daou and Antoine Maignan and Philippa H. McGuinness and Markus König and Helge Rosner and Seo Jin Kim and Seunghyun Khim and Romain Grasset and Marcin Konczykowski and Evyatar Tulipman and Juan Felipe Mendez-Valderrama and Debanjan Chowdhury and Erez Berg and Andrew P. Mackenzie},
   doi = {10.1073/pnas.2307334120},
   issn = {10916490},
   journal = {Proceedings of the National Academy of Sciences of the United States of America},
   pages = {e2307334120},
   pmid = {37639594},
   title = {Investigation of {P}lanckian behavior in a high-conductivity oxide: \ce{PdCrO2}},
   volume = {120},
   year = {2023}
}

@article{Zhang2024,
   author = {Yi Zhang and Fred Tutt and Guy N. Evans and Prachi Sharma and Greg Haugstad and Ben Kaiser and Justin Ramberger and Samuel Bayliff and Yu Tao and Mike Manno and Javier Garcia-Barriocanal and Vipul Chaturvedi and Rafael M. Fernandes and Turan Birol and William E. Seyfried and Chris Leighton},
   doi = {10.1038/s41467-024-45239-6},
   issn = {20411723},
   journal = {Nature Communications},
   pages = {1399},
   publisher = {Springer US},
   title = {Crystal-chemical origins of the ultrahigh conductivity of metallic delafossites},
   volume = {15},
   year = {2024}
}

@book{Ziman1964,
   author = {J M Ziman},
   title = {Principles of the theory of solids},
   year = {1964}
}

@article{Zitzlsperger2003,
   author = {M Zitzlsperger and R Onderka and M Suhrke and U Rössler and Weiss and W Wegscheider and M Bichler and R Winkler and Y Hirayama and K Muraki},
   doi = {10.1209/epl/i2003-00185-0},
   journal = {Europhysics Letters},
   pages = {382},
   title = {Chaos and open orbits in hole-antidot arrays with non-isotropic {F}ermi surface},
   volume = {61},
   year = {2003}
}

@article{Oka2019,
   author = {Takushi Oka and Shingo Tajima and Ryoya Ebisuoka and Taiki Hirahara and Kenji Watanabe and Takashi Taniguchi and Ryuta Yagi},
   doi = {10.1103/PhysRevB.99.035440},
   journal = {Physical Review B},
   pages = {035449},
   publisher = {American Physical Society},
   title = {Ballistic transport experiment detects {F}ermi surface anisotropy of graphene},
   volume = {99},
   year = {2019}
}

@article{Beenakker1991,
   author = {C. W.J. Beenakker and H. van Houten},
   doi = {10.1016/S0081-1947(08)60091-0},
   issn = {00811947},
   journal = {Solid State Physics},
   pages = {1},
   title = {Quantum Transport in Semiconductor Nanostructures},
   volume = {44},
   year = {1991}
}

@article{Daughton1999,
   author = {J M Daughton},
   doi = {10.1016/S0304-8853(98)00376-X},
   journal = {Journal of Magnetism and Magnetic Materials},
   pages = {334-342},
   title = {{GMR} applications},
   volume = {192},
   year = {1999}
}

@article{Kumar2022,
   author = {Sushant Kumar and Christian Multunas and Benjamin Defay and Daniel Gall and Ravishankar Sundararaman},
   doi = {10.1103/PhysRevMaterials.6.085002},
   journal = {Physical Review Materials},
   pages = {085002},
   publisher = {American Physical Society},
   title = {Ultralow electron-surface scattering in nanoscale metals leveraging {F}ermi-surface anisotropy},
   volume = {6},
   year = {2022}
}

@article{Mazzola2022,
   author = {F. Mazzola and C. M. Yim and V. Sunko and S. Khim and P. Kushwaha and O. J. Clark and L. Bawden and I. Marković and D. Chakraborti and T. K. Kim and M. Hoesch and A. P. Mackenzie and P. Wahl and P. D.C. King},
   doi = {10.1038/s41535-022-00428-8},
   journal = {npj Quantum Materials},
   pages = {20},
   publisher = {Nature Research},
   title = {Tuneable electron–magnon coupling of ferromagnetic surface states in \ce{PdCoO2}},
   volume = {7},
   year = {2022}
}

@article{Seo2023,
   author = {Dongmin Seo and Gihyeon Ahn and Gaurab Rimal and Seunghyun Khim and Suk Bum Chung and A. P. Mackenzie and Seongshik Oh and S. J. Moon and Eunjip Choi},
   doi = {10.1038/s41535-023-00607-1},
   journal = {npj Quantum Materials},
   pages = {74},
   publisher = {Nature Research},
   title = {Interaction of in-plane Drude carrier with c-axis phonon in PdCoO2},
   volume = {8},
   year = {2023}
}

@article{Yao2024,
   author = {Xiaoping Yao and Yechen Xun and Ziye Zhu and Shu Zhao and Wenbin Li},
   doi = {10.1103/PhysRevB.109.075110},
   journal = {Physical Review B},
   pages = {075110},
   publisher = {American Physical Society},
   title = {Origin of the high electrical conductivity of the delafossite metal PdCoO2},
   volume = {109},
   year = {2024}
}

\end{document}